# Optimization design and analysis for the mechanical test platform of scientific probe module of the Cool Planet Imaging Coronagraph

Lingyi Kong[a], Jiangpei Dou[a], Wei Guo[a], Mingming Xu[*a], Shu Jiang[a] and Bo Chen[a]
[a] Solar and Space Instrument Research Laboratory, Nanjing Institute of Astronomical Optics & Technology, Chinese Academy of Sciences, Nanjing, Jiangsu Province, China;

**Abstract**: This paper optimizes the design and analysis of the mechanical test platform for the scientific probe module of the Cool Planet Imaging Coronagraph, which is the fifth part of the China Space Station survey Telescope. First, according to the module layout and economic requirements, the preliminary structural design of the module mechanical test platform is carried out, and the stiffness sensitive parameters of the assembly are identified to determine the optimization parameters. The Central Composite Design method is used to design the test platform, and a third-order regression model is constructed for response surface analysis. The third-order response surface model of the fundamental frequency and amplitude of the test platform is obtained by fitting the test data with the least squares method, and the structure of the module mechanical test platform is determined. The modal analysis is carried out to determine the fundamental frequency and vibration modes of the mechanical test platform. The vibration response of the platform is simulated by sine, random and swept frequency vibration simulations. The response surface fitting algorithm is verified by the test platform swept frequency test. The agreement between the response surface fitting algorithm and the experiment is good. The fundamental frequency of the test platform is 436.2Hz (＞300Hz), which meets the design index requirements of the test platform and can accurately guide the optimization design work. At the same time, it provides the theoretical basis and design method for the structural design of the Chinese manned space station.

---
* Corresponding Author, Mingming Xu, No.188 Bancang Street, Xuanwu District, Nanjing, Jiangsu Province, China; E-mail: mingxu@niaot.ac.cn.

## 0 Introduction

The China Space Station survey Telescope (CSST) is the first large-scale space survey telescope independently developed by China. It will carry out cutting-edge scientific research in the formation and evolution of cosmic structure, dark matter and dark energy, exoplanets and solar system bodies. Among them, the Cool Planet Imaging Coronagraph (CPI-C) is one of the important observation equipment of CSST. It is responsible for the high-contrast imaging detection of exoplanets, which has very important scientific significance for my country's independent exploration of exoplanets and the search for the "second Earth" [1].

The CPI-C is divided into a scientific probe module and an integrated power distribution module. The scientific probe module is the optical imaging module in the CPI-C. It integrates an internal occultation optical path system. By adjusting the internal occultation structure, it can block or eliminate starlight and capture and detect planetary light. It is a key technology for confirming terrestrial planets, but it also faces unprecedented challenges. The contrast between a planet and its host star is very different. The target imaging contrast of the scientific probe module needs to reach 10-8 to achieve high-contrast direct imaging research on exoplanets [2-6]. The scientific probe module integrates sensitive optical components such as the internal optical path system and visible light imaging cameras and wavefront cameras. As an important optical system module, the overall rigidity, strength and structural stability of the module are the basis for achieving the functional performance of the module. As a space optical payload, the scientific probe module must meet the strict mechanical environment requirements of CSST during design to ensure the structural stability and safety of the module under the dynamic environment conditions during rocket launch [7]. Therefore, after the module is assembled, a mechanical environment test is required to investigate the internal sensitive components and single-machine response, and further revise the module design and mechanical environment conditions. As the intermediate component between the vibration table and the module, the stiffness and mechanical transfer characteristics of the test platform are particularly important for the safety of mechanical tests and the accuracy of test results. Therefore, the development and design of the mechanical test platform is the basis for ensuring the normal launch of the module and subsequent scientific detection.

This paper conducts a preliminary design of the module mechanical test platform, uses finite element software to optimize the design and sensitivity analysis of the mechanical test platform, determines the final design scheme and conducts mechanical environment vibration analysis, and finally conducts mechanical environment experimental tests on the design scheme to provide a technical basis for the scientific probe module of the space telescope coronagraph.

## 1 Structural features of the test platform

The scientific probe module has a flat structure,

with three upper interfaces A, B, and C distributed on the three end faces of the module, as shown in Figure 1.

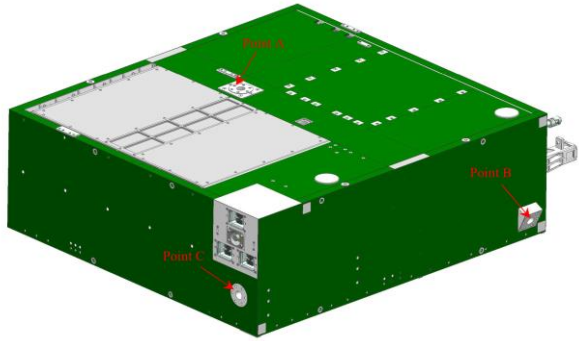

Figure 1 Structural model of scientific probe module

The design of the test platform needs to consider economy, module installation accessibility, platform stiffness and interface dynamic response. The test platform is mainly divided into the platform base plate and the interface mounting brackets. In order to reduce the module installation height, the module is installed horizontally on the test platform during design. The mounting bracket at point A is an integrated machined boss structure, and the mounting brackets at points B and C of the interface are assembled by screwing after plate processing, which reduces the cost of materials and machining. At the same time, considering the selection of the subsequent test vibration table, the lightweight aluminum alloy 2A12-T4 with high specific stiffness is selected as the test platform material during design to reduce the mass of the test platform and reduce the thrust index requirements of the vibration test table. The physical properties of the aluminum alloy 2A12-T4 material are shown in Table 1.

**Tab. 1** Physical Properties of 2A12-T4

| density | Elastic modulus | Poisson's ratio |
|---|---|---|
| 2.8e3kg/m^3 | 71GPa | 0.33 |

According to the structural characteristics of the scientific probe module, the test platform was preliminarily designed, and the structural diagram is shown in Figure 2. The mounting bracket at point A has good rigidity and low response amplification, so this paper mainly optimizes and analyzes the mounting bracket structure at interfaces B and C. Since the bracket is made of plate processing and splicing, the factors affecting the overall rigidity of the test platform and the module interface response are mainly the thickness and position of the back rib plate. Affected by the spatial layout, the optimization parameters are the thickness parameters of the rib plates on both sides of the mounting brackets at interfaces B and C, whose structures are shown in Figure 3. According to the design requirements, the fundamental frequency of the module mechanical test platform is 3 to 5 times the fundamental frequency of the module [8-9]. For the mechanical test platform of the scientific probe module, its fundamental frequency needs to be greater than 300Hz, and the vibration response value should be low.

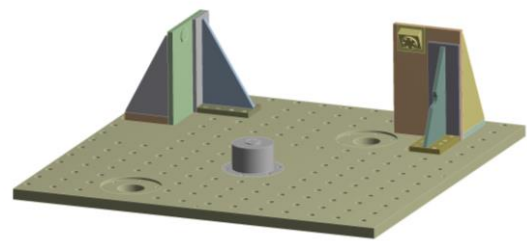

Figure 2 Schematic diagram of mechanical test platform

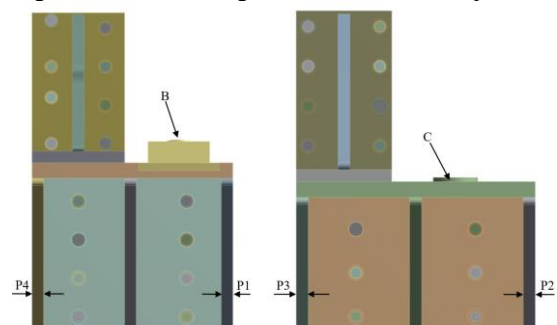

Figure 3 Parameter diagram of the BC point mounting bracket

## 2 Modal analysis of the test platform

The objective of modal analysis is to determine the product's natural frequencies and mode shapes, requiring the test platform to exhibit high stiffness with minimal response while preventing resonance between the platform and test module to ensure testing safety.

According to vibration theory and finite element modal analysis theory, the general dynamic equation of structural vibration of elastic vibration systems with multiple degrees of freedom is [10-11]:

$$M\{\ddot{x}\} + C\{\dot{x}\} + K\{x\} = F(t) \quad (1)$$

Where $M$ is the mass matrix; $C$ is the damping matrix; $K$ is the stiffness matrix; $F(t)$ is the excitation force column vector; $\{\ddot{x}\}$ represents the acceleration response; $\{\dot{x}\}$ represents the velocity response; $\{x\}$ represents the displacement response.

When solving the fundamental frequency and vibration mode of the free vibration of the metal assembly structure, the structural damping is small and has a very small impact on it, which can be ignored. Since there is no force in the modal analysis, that is $F(t) = \{0\}$, equation (1) can be simplified to the undamped structural vibration motion equation:

$$M\{\ddot{x}\} + K\{x\} = 0 \quad (2)$$

Then its characteristic equation is

$$(K - \omega^2 M)M = \{0\} \quad (3)$$

Where: $\omega$ represents the fundamental frequency of the system.

## 3 Optimization design of test platform model

### 3.1 Experimental Design

The test platform was designed for the experiment. In order to study the influence of each rib on the result, the value ranges of the right rib thickness $p_1$, the left rib thickness $p_4$, and the right rib thickness $p_2$, the left rib thickness $p_3$ of the B-interface mounting bracket, and the C-interface mounting bracket were set, as shown in Figure 3 and Table 2.

**Tab. 2** Value range

| range | $p_1$/mm | $p_2$/mm | $p_3$/mm | $p_4$/mm |
|---|---|---|---|---|
| Minimum | 2 | 2 | 2 | 2 |
| Maximum | 20 | 20 | 20 | 20 |

To simplify the calculation, the test platform simulation model only retains the screw connection between the B and C interface mounting brackets and the base plate. The screw size is M8, the screw preload is set to 10750N, the contact relationship is friction contact, the friction coefficient is 0.1, and the contact relationship between the remaining parts is simplified to binding contact. The Central Composite Design method is used to design the test platform, and the fundamental frequency $f$ and amplitude $A$ results obtained by prestressed modal simulation are shown in Table 3. Figure 4 shows the sensitivity distribution of the four parameters.

**Tab. 3** Experimental scheme and results

| Sample points | $p_1$ mm | $p_2$ mm | $p_3$ mm | $p_4$ mm | $f$ Hz | $A$ mm |
|---|---|---|---|---|---|---|
| 1 | 11 | 11 | 11 | 11 | 490.387 | 25.427 |
| 2 | 2 | 11 | 11 | 11 | 176.557 | 201.983 |
| 3 | 6.5 | 11 | 11 | 11 | 482.45 | 49.0672 |
| 4 | 20 | 11 | 11 | 11 | 499.335 | 23.297 |
| 5 | 15.5 | 11 | 11 | 11 | 491.588 | 24.409 |
| 6 | 11 | 2 | 11 | 11 | 248.958 | 240.044 |
| 7 | 11 | 6.5 | 11 | 11 | 490.348 | 25.427 |
| 8 | 11 | 20 | 11 | 11 | 490.387 | 25.427 |
| 9 | 11 | 15.5 | 11 | 11 | 490.388 | 25.427 |

| Sample points | $p_1$ mm | $p_2$ mm | $p_3$ mm | $p_4$ mm | $f$ Hz | $A$ mm |
|---|---|---|---|---|---|---|
| 10 | 11 | 11 | 2 | 11 | 176.279 | 201.753 |
| 11 | 11 | 11 | 6.5 | 11 | 483.356 | 51.274 |
| 12 | 11 | 11 | 20 | 11 | 488.078 | 24.945 |
| 13 | 11 | 11 | 15.5 | 11 | 489.011 | 25.227 |
| 14 | 11 | 11 | 11 | 2 | 250.352 | 240.687 |
| 15 | 11 | 11 | 11 | 6.5 | 490.388 | 25.427 |
| 16 | 11 | 11 | 11 | 20 | 490.387 | 25.427 |
| 17 | 11 | 11 | 11 | 15.5 | 490.394 | 25.425 |
| 18 | 2 | 2 | 2 | 2 | 176.269 | 200.071 |
| 19 | 6.5 | 6.5 | 6.5 | 6.5 | 477.724 | 44.087 |
| 20 | 20 | 2 | 2 | 2 | 176.301 | 201.754 |
| 21 | 15.5 | 6.5 | 6.5 | 6.5 | 483.978 | 51.355 |
| 22 | 2 | 20 | 2 | 2 | 176.269 | 200.073 |
| 23 | 6.5 | 15.5 | 6.5 | 6.5 | 477.721 | 44.086 |
| 24 | 20 | 20 | 2 | 2 | 176.301 | 201.754 |
| 25 | 15.5 | 15.5 | 6.5 | 6.5 | 484.291 | 51.668 |
| 26 | 2 | 2 | 20 | 2 | 176.588 | 201.98 |
| 27 | 6.5 | 6.5 | 15.5 | 6.5 | 481.56 | 48.348 |
| 28 | 20 | 2 | 20 | 2 | 248.954 | 239.883 |
| 29 | 15.5 | 6.5 | 15.5 | 6.5 | 490.116 | 24.229 |
| 30 | 2 | 20 | 20 | 2 | 176.588 | 201.98 |
| 31 | 6.5 | 15.5 | 15.5 | 6.5 | 481.526 | 48.327 |
| 32 | 20 | 20 | 20 | 2 | 250.361 | 240.688 |
| 33 | 15.5 | 15.5 | 15.5 | 6.5 | 490.118 | 24.229 |
| 34 | 2 | 2 | 2 | 20 | 176.27 | 200.079 |
| 35 | 6.5 | 6.5 | 6.5 | 15.5 | 477.723 | 44.087 |
| 36 | 20 | 2 | 2 | 20 | 176.301 | 201.754 |
| 37 | 15.5 | 6.5 | 6.5 | 15.5 | 484.314 | 51.688 |
| 38 | 2 | 20 | 2 | 24 | 176.27 | 200.079 |
| 39 | 6.5 | 15.5 | 6.5 | 15.5 | 477.718 | 44.085 |
| 40 | 20 | 20 | 2 | 20 | 176.301 | 201.754 |
| 41 | 15.5 | 15.5 | 6.5 | 15.5 | 484.292 | 51.669 |
| 42 | 2 | 2 | 20 | 20 | 176.587 | 201.98 |
| 43 | 6.5 | 6.5 | 15.5 | 15.5 | 481.523 | 48.32 |
| 44 | 20 | 2 | 20 | 20 | 248.962 | 240.046 |
| 45 | 15.5 | 6.5 | 15.5 | 15.5 | 490.149 | 24.229 |
| 46 | 2 | 20 | 20 | 20 | 176.587 | 201.98 |
| 47 | 6.5 | 15.5 | 15.5 | 15.5 | 481.552 | 48.342 |
| 48 | 20 | 20 | 20 | 20 | 496.712 | 22.901 |
| 49 | 15.5 | 15.5 | 15.5 | 15.5 | 490.149 | 24.229 |

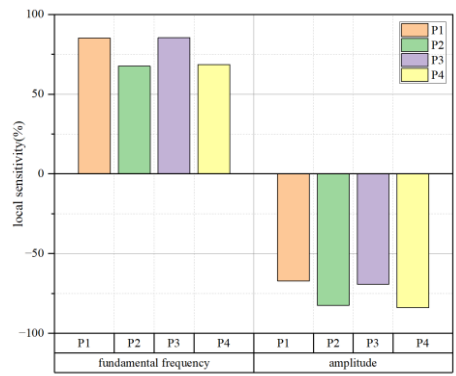

Figure 4 Sensitivity distribution diagram

Comprehensive model layout and test results show that $p_1 \sim p_4$ is proportional to the fundamental frequency and inversely proportional to the amplitude. When a parameter value in $p_1 \sim p_4$ is lower than a certain threshold (<6.5mm), the fundamental frequency presents a local mode of the single board; when it is higher than a certain threshold, the maximum change of the fundamental frequency is 4.3%, and the maximum change of the amplitude is 10%, both of which are relatively low.

From the sensitivity distribution in Figure 4, $p_1$、$p_3$ can be seen that the absolute values of the contribution of and to the fundamental frequency and amplitude are both greater than 80%, $p_2$、$p_4$ the absolute values of the contribution of and to the fundamental frequency and amplitude are both less than 70%, and the correlation between the parameters is relatively large.

### 3.2 Response surface model establishment and analysis

The response surface method enables multi-parameter-influenced target optimization by generating response-design variable correlations through experimental data fitting. Xing et al. [12-15] analyzed heat sink structural/flow parameter effects on thermal dissipation efficiency, employing RSM to develop quadratic regression models and identify Pareto-optimal solutions that enhanced heat exchanger efficiency by 12.7–18.3%. Wang [16] conducted RSM-based analysis of gas-liquid cyclone separators, quantifying structural/operational parameter impacts on separation efficiency and pressure drop, establishing high-efficiency low-resistance separation systems. Liu et al. [17] applied RSM to space gravitational wave detection satellites' ultra-stable structures, quantifying elastic modulus/support layout effects on anti-interference capability (error <8%). Qin [18] integrated CFD with RSM for G1 micromixer optimization, constructing quadratic polynomial models with parallel PSO to increase flow rate by 13.51% and mixing efficiency by 2.45%. Ge [19] utilized RSM for electric vibration table optimization, enhancing vibration uniformity/load capacity by 15.2–22.6% through parameter interaction analysis. Zhang [20] optimized high-temperature strain gauge grids via RSM, reducing measurement errors by 22.7% and temperature drift coefficients by 18.5%. Liu et al. [21-22] implemented RSM for structural-acoustic radiation optimization, achieving 5.8–9.3 dB noise reduction. In this section, the response surface method is used to study the relationship between the stiffener wall thickness and the fundamental frequency and amplitude of the test platform. The functional relationship between the parameters is shown in formula (4).

$$y_f = f(x), \quad y_A = g(x) \quad (4)$$

Where $y_f$、$y_A$ are the fundamental frequency and amplitude of the test platform, respectively.

The actual functional relationship between the response $y$ and the variable $x = (p_1, p_2, p_3, p_4)$ is very complex, so an approximate functional relationship is used to fit a regression equation containing a cubic term to obtain a third-order response surface model, whose expression is shown in formula (5):

$$y = \sum_{i=1}^{k} a_i x_i + \sum_{i=1}^{k} a_{iii} x_i^2 + \sum_{i=1}^{k} a_{iii} x_i^3 + \sum_{i=1,j=1}^{k} a_{ij} x_i x_j + \varepsilon \quad (5)$$

In the formula, $a$ represents the regression coefficient, $\varepsilon$ is the deviation, $k$ and is the number of variables.

In order to prove the fitting accuracy of the fitting model, it is necessary to perform error analysis on the fitting model. Therefore, the following parameters are introduced: $sst$ is the total sum of squares, that is, the sum of squares of the difference between the test value and the average value of the test value; $ssr$ is the regression sum of squares, that is, the sum of squares of the difference between the fitting value and the average value of the test value; $sse$ is the residual sum of squares, that is, the sum of squares of the difference between the test value and the fitting value [23]. The relationship between the parameters is shown in formula (6).

$$sst = \sum_{i=1}^{n} (y_i - \bar{y})^2$$
$$ssr = \sum_{i=1}^{n} (y_a - \bar{y})^2$$
$$sst = ssr + sse \quad (6)$$

Therefore, the error determination coefficient of the fitting model is shown in formula (7).

$$r^2 = \frac{ssr}{sst} = 1 - \frac{sse}{sst}$$
$$r_{adj}^2 = 1 - \frac{sse/(n-d)}{sst/(n-1)}$$
(7)

Among them, $r^2$ is the accuracy determination coefficient, $r_{adj}^2$ is the correction coefficient for removing the number of terms in the regression equation, $n$ is the total number of experimental values, and $d$ is the total number of terms in the fitting equation. When the fitting model accuracy is higher, $r^2$ is closer to 1, $r^2$ and $r_{adj}^2$ is closer to.

Based on the above model and the data in Table 3, the fitting is performed, and the local modal sample points with large discreteness are eliminated during the fitting, and the fundamental frequency response surface model of the test platform is obtained as follows:

$$y_f = 0.0254p_1^3 - 0.0014p_2^3 + 0.016p_3^3 - 0.016p_4^3$$
$$- 0.997p_1^2 + 0.0535p_2^2 - 0.727p_3^2$$
$$+ 0.0607p_4^2 - 0.00185p_1p_2$$
$$+ 0.023p_1p_3 - 0.0015p_1p_4$$
$$- 0.0037p_2p_3 - 0.0037p_2p_4$$
$$- 0.0037p_3p_4 + 12.834p_1 - 0.53p_2$$
$$+ 10.25p_3 - 0.62p_4 + 391.7$$
(8)

The accuracy determination coefficient $r^2$ is 0.9976, and the correction coefficient $r_{adj}^2$ is 0.9974. The two are almost consistent and both tend to 1. The fitting model is effective and of excellent quality.

Similarly, the amplitude response surface model of the test platform is obtained by eliminating the sample points with large discreteness as follows:

$$y_A = -0.04p_1^3 + 0.0084p_3^3 + 1.886p_1^2 - 0.393p_3^2$$
$$- 28.62p_1 + 5.89p_3 + 137.58$$
(9)

The accuracy determination coefficient $r^2$ is 0.9999, and the correction coefficient $r_{adj}^2$ is 0.9999. The two are almost consistent and both tend to 1. The fitting model is effective and of excellent quality.

Since the ABC interface of the test platform is the product installation position, when the module-level mechanical environment test is performed, and $p_1$、$p_2$ have a greater impact on the stiffness of the assembly. At the same time, it can be observed from Table 3 that the $p_3$ threshold is around 11mm and the $p_4$ threshold is around 6.5mm. When the actual test platform is assembled, the ribs are fastened with M6 screws. Considering the material of the test platform and the safety of disassembly and assembly, M8-M6 wire screw sleeves need to be nested on the ribs. Therefore, the final thickness of each rib should not be less than 10mm. Based on the above results, it can be preliminarily inferred that the value of $p_3$ is an integer and the value of $p_4$ is 10mm.

When $p_3$、$p_4$ are both 10mm, formula (8) can be simplified to formula (10):

$$y_{f1} = 0.0254p_1^3 - 0.0014p_2^3 - 0.997p_1^2 + 0.0535p_2^2$$
$$- 0.00185p_1p_2 + 13.049p_1 - 0.604p_2$$
$$+ 435.4$$
(10)

When $p_3$ is 11mm and $p_4$ is 10mm, formula (8) can be simplified to formula (11):

$$y_{f2} = 0.0254p_1^3 - 0.0014p_2^3 - 0.997p_1^2 + 0.0535p_2^2$$
$$- 0.00185p_1p_2 + 13.349p_1 - 0.608p_2$$
$$+ 435.64$$
(11)

The fundamental frequency response surface cloud diagrams of the test platform under the two design conditions are shown in Figures 5 and 6. At the same time, the output amplitude response surface cloud diagram is shown in Figure 7:

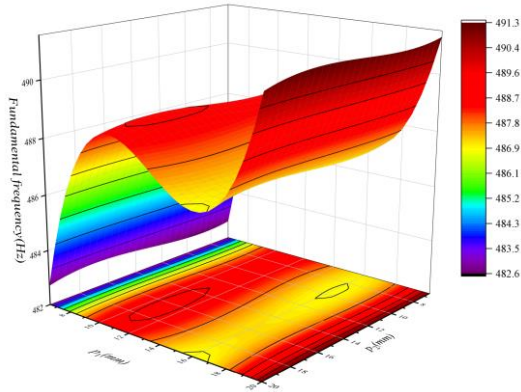

Figure 5 $p_3$=10mm base frequency response surface cloud diagram

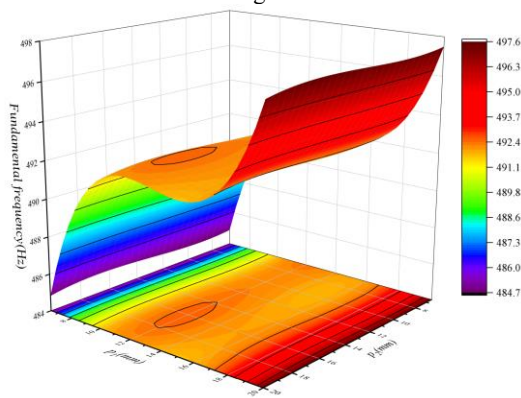

Figure 6 $p_3$=11mm fundamental frequency response surface cloud diagram

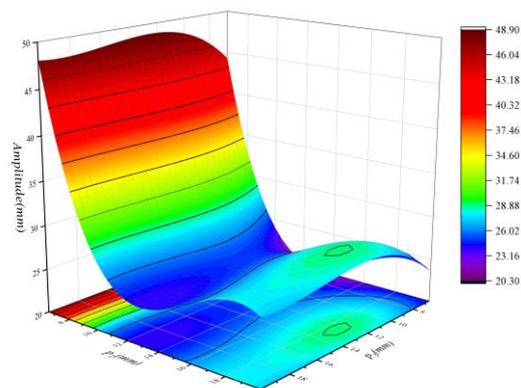

Figure 7 Amplitude response surface cloud diagram

As shown in Figures 5 and 6, the fundamental frequency is slightly higher when is 11mm, but only 0.6% higher; since the values of each parameter are integers, this paper does not optimize the optimal value. Considering the influence of fundamental frequency, amplitude and economy, the optimal design values of $p_1$ and $p_2$ are 12mm and 10mm respectively. Therefore, the four parameters of the test platform are 12mm, 10mm, 10mm, and 10mm respectively, and the fitting fundamental frequency and amplitude of the test platform are 489.98Hz and 24.6mm respectively.

## 4 Dynamics simulation analysis
### 4.1 Modal analysis

Utilizing the aforementioned parameters, the test platform structure was geometrically refined. An automatic meshing scheme employing SOLID186 and SOLID187 elements generated 897,131 elements and 1,846,900 nodes (average mesh quality: 0.829) as shown in Figure 8. Subsequently, a prestressed modal analysis was performed under the boundary conditions defined in Section 2 (including a fully fixed constraint on the baseplate's inferior surface) extracted the first 32 structural modes, with the mode shapes of the first to six-order shown in Figures 9 to Figures 14.

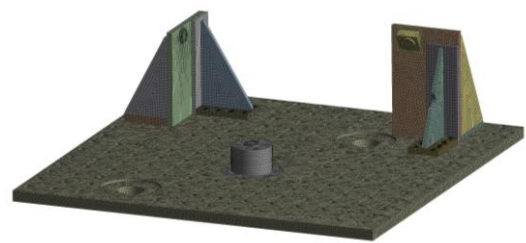

Figure 8 Grid model

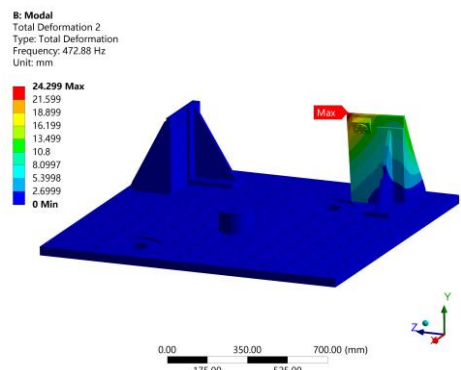

Figure 9 First-order vibration mode

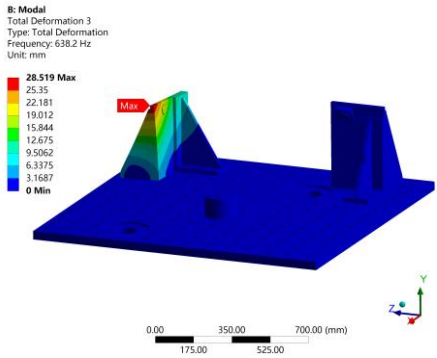
Figure 10 Second-order vibration mode

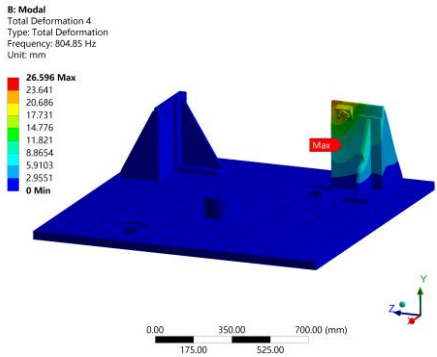
Figure 11 Third-order vibration mode

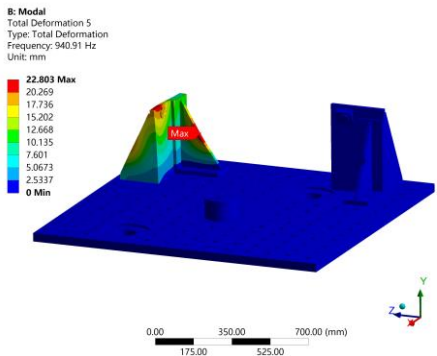
Figure 12 Fourth-order vibration mode

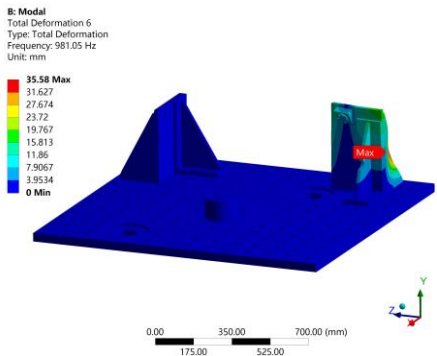
Figure 13 Fifth-order vibration mode

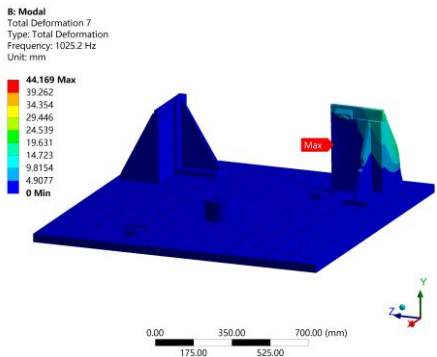
Figure 14 Sixth-order vibration mode

As can be seen from the results in Table 4, the first to sixth-order natural frequencies of the test platform are 472.88 Hz, 638.2 Hz, 804.85 Hz, 940.9 Hz, 981.05 Hz, and 1025.2 Hz, respectively. Due to model refinement, minor discrepancies exist between the fundamental frequency, amplitudes, and the fitting results. The mass fractions for each mode in various directions are presented in the table below. Based on the mass fractions, the mode shapes primarily involve translational motion of the part along the X and Z directions, rotation about the Y direction, and coupled vibrations in these directions. The amplitudes in all other directions are negligible.

**Tab. 4** Comparison of results

| Order | $f$/Hz | Mass fraction (%) | | | | | |
|---|---|---|---|---|---|---|---|
| | | X | Y | Z | RX | RY | RZ |
| first | 472.88 | 0.595 | 0.049 | 2.41 | 0.037 | 1.37 | 0.033 |
| second | 638.2 | 1.69 | 0.025 | 0.794 | 0.04 | 1.8 | 0.007 |
| third | 804.85 | 2.23 | 0.013 | 1.57 | 0.007 | 0.024 | 0.08 |
| fourth | 940.9 | 0.462 | 0.014 | 0.692 | 0.023 | 4.6 | 0.012 |
| fifth | 981.05 | 0.808 | 0.021 | 0.199 | 0.006 | 0.024 | 0.059 |
| sixth | 1025.2 | 0.106 | 0 | 0.024 | 0 | 0.015 | 0.008 |

### 4.2 Vibration simulation analysis

In order to ensure the structural safety of the coronagraph scientific probe module during the dynamic test, it is necessary to simulate and analyze the dynamic characteristics of the test platform. The sinusoidal and random vibration test conditions are shown in Table 5 and 6.

**Tab. 5** Sinusoidal vibration test conditions

| X direction(Sweep rate:2oct/min) | | | | |
|---|---|---|---|---|
| $f$/Hz | 5~10 | 10~14 | 14~75 | 75~100 |
| $A$ | 7.45mm | 3g | 5.5g | 4g |
| Y、Z direction(Sweep rate:2oct/min) | | | | |
| $f$/Hz | 5~10 | 10~14 | 14~70 | 70~100 |
| *A* | 9.93mm | 4g | 7g | 4g |

**Tab. 6 Random vibrations test conditions**

| | X、Y、Z direction | |
|---|---|---|
| PSD | 20~100Hz | 3dB/oct |
| | 100~600Hz | 0.02g2/Hz |
| | 600~2000Hz | -9dB/oct |
| RMS | 4.02g | |
| Duration | 3min | |

Based on the aforementioned vibration conditions, simulation analysis was performed on the test platform, with the structural damping coefficient referenced from Section 5. The maximum stress of the sinusoidal vibration response was 1.75MPa in the Z direction; for the random vibration response, the calculated 3σ peak stress level in the Z direction was 26.485MPa, obtained using modal superposition-based analysis with the input PSD excitations. The test platform is made of 2A12-T4 aluminum alloy, whose allowable stress is 210MPa. Based on this 3σ random stress level, the minimum safety factor of the test platform is 7.93, which is greater than the structural component yield limit safety factor of 1.2 and the failure load safety factor of 1.35 required by aerospace [24]. Therefore, the test platform meets the structural safety design requirements.

The test platform was simulated by sweeping the frequency in three directions in the range of 0~500Hz with an input of 0.2g, and the acceleration response of the test platform under the fundamental frequency condition was obtained. Among them, the test platform is insensitive to Y direction excitation, so only the acceleration response curves of the path of B point (the center of B point is connected with the bottom, see Figure 15 for details) and the paths near the Z direction at the fundamental frequency in the X and Z directions are output as shown below.

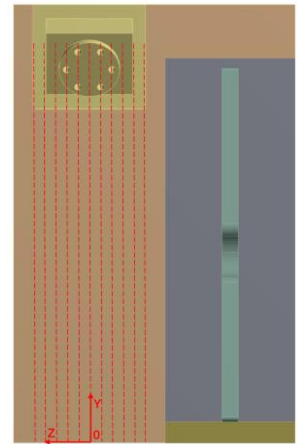

Figure 15 B point Mounting Path Diagram

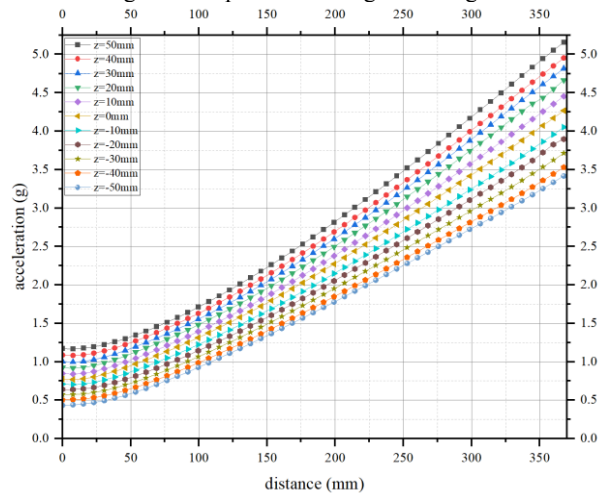

Figure 16 Acceleration response curve of X direction

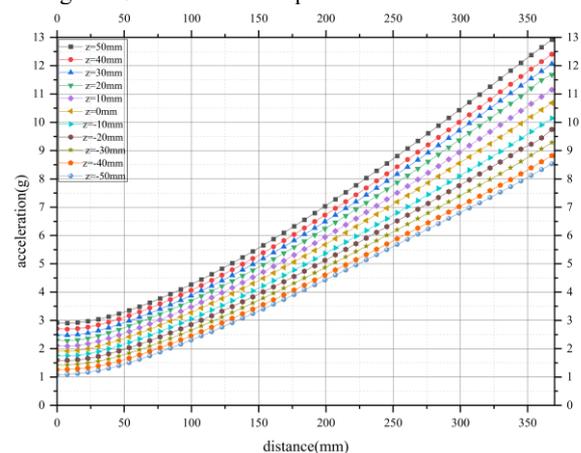

Figure 17 Acceleration response curve of Z direction

From the fundamental frequency vibration mode and the path response diagram, it is evident that resonant responses occur in both the X and Z directions at the fundamental frequency, with the response in the Z direction being slightly higher. The curve exhibits a steeper gradient along the path from the bottom of the mounting base to the center height of mounting B point. When sweeping in the X direction the acceleration response at the center height of mounting B point is amplified by approximately 25.8 times, and the acceleration response in the Z direction is amplified by approximately 53.4 times. Concurrently, the average response increment per 10 mm interval is approximately 0.44 g. Each curve, spanning a distance of 100 mm to 360 mm from the bottom of the mounting base to the center of mounting B point, is essentially linear. Therefore, if further response optimization is undertaken in the future, incorporating transverse ribs between the central and right rib plates of mounting base B can be designed to reduce the response at the interface.

## 5 Experimental verification

To verify the accuracy of the fitting and simulation results, the test platform was mounted on an 18-ton vibration table for 0.2g sinusoidal sweep testing across 10–2000 Hz in three orthogonal directions (Figure 18). Accelerometer monitoring points were installed near interfaces A, B, and C to record response data. The response curves of the 0.2g sweep frequency simulation and experimental tests are shown in Figures 19 to 24.

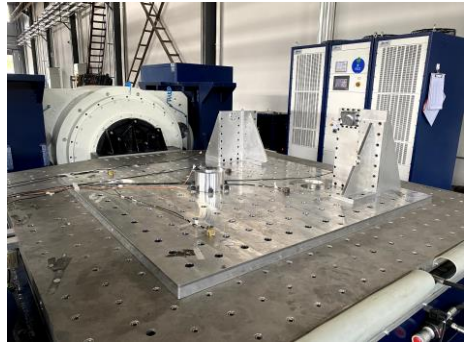

Figure 18 Test platform frequency sweep test

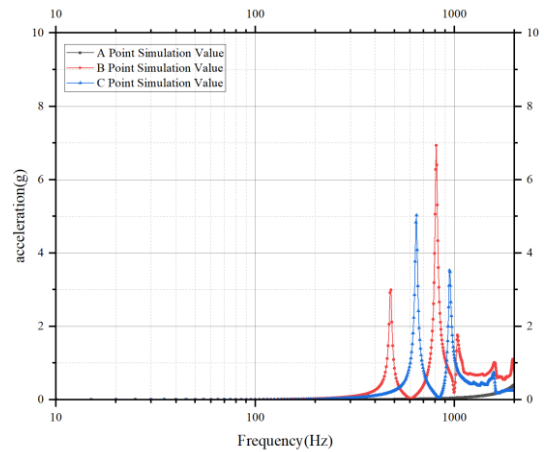

Figure 19 X direction acceleration simulation curve

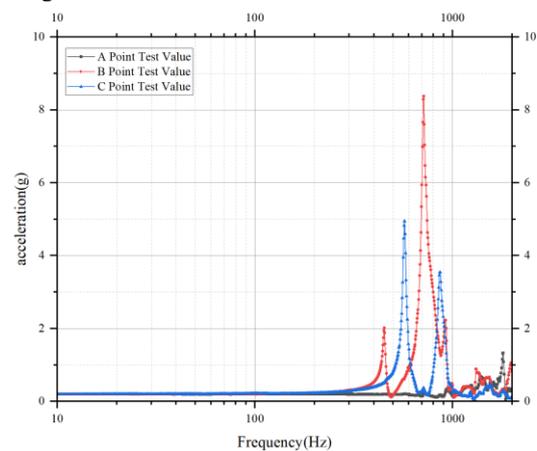

Figure 20 X direction acceleration test curve

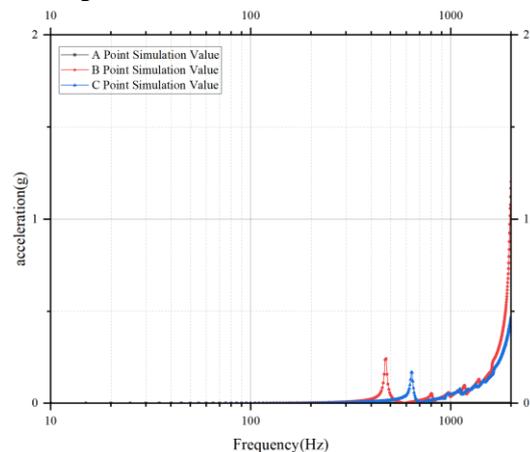

Figure 21 Y direction acceleration simulation curve

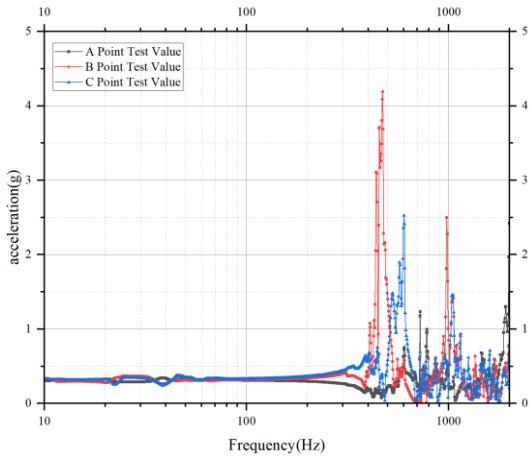
Figure 22 Y direction acceleration test curve

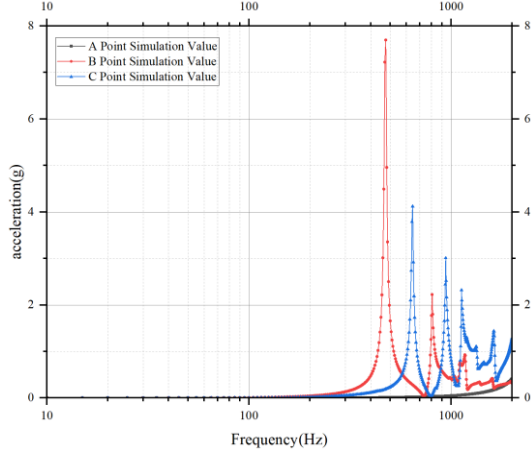
Figure 23 Z direction acceleration simulation curve

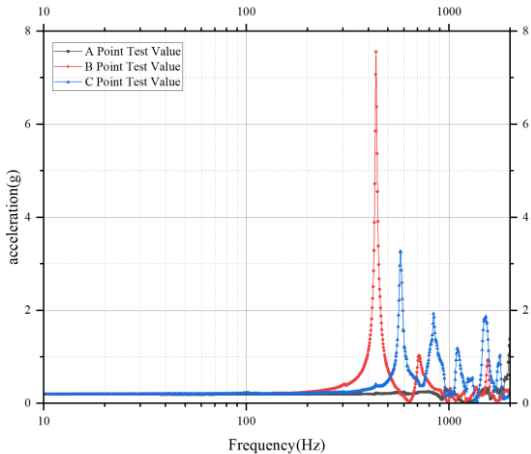
Figure 24 Z direction acceleration test curve

Given minimal modal coupling in resonance peaks and the platform's straightforward configuration, the half-power bandwidth method was employed to extract structural damping ratios from experimental curves. For simulations, damping values in the X and Z directions were set to 0.014 and 0.011 respectively, aligning with experimental measurements. The irregular Y direction experimental curves precluded precise damping evaluation; thus, the simulated structural damping for this direction was assigned 0.012 based on comprehensive analysis of other directional data and partial Y direction resonance peaks.

From the analysis of the curve results, it can be seen that the resonant frequencies of the test platform in the X direction are 450Hz, 571Hz, 715Hz, and 865Hz; the resonant frequencies in the Y direction are 474Hz, 604Hz, 984Hz, and 1053Hz; and the resonant frequencies in the Z direction are 436.2Hz, 577Hz, and 837Hz, which are compared with the simulation results as shown in Table 7. The first-order, fifth-order, and sixth-order results have good consistency, while the second-order to fourth-order results have certain divergences.

**Tab. 7** Comparison of results

| Order | Simulation Results | Test results | Average Error |
|---|---|---|---|
| first | 472.88 | 436.2/450/474 | 4.2% |
| second | 638.2 | 571/577/604 | 8.5% |
| third | 804.85 | 715 | 11.1% |
| fourth | 940.9 | 837/865 | 9.5% |
| fifth | 981.05 | 984 | 0.3% |
| sixth | 1025.2 | 1053 | 0.07% |
| Average total error | | | 5.6% |

Experimental and simulated curves exhibit substantial agreement along the X and Z directions, with only minor deviations observed in select resonance peak magnitudes, thereby validating the simulation's effectiveness. However, Y direction experimental curves exhibit significant anomalies and substantial divergence from simulation results. Theoretically, modal shapes, effective mass participation, and structural configuration indicate that the Y direction possesses the highest stiffness among all three orthogonal directions. Consequently, prominent resonance peaks should not manifest in lower-order modes. The possible reasons for the divergence are as follows:

1) Simulation condition: The simulation only considered the bolted connection and fastening relationship between Mounts B and C and the platform base plate. The bolted connections between the platform base plate and the test bench, as well as the mount plates, were neglected. Additionally, subject to site constraints, only partial fastening screws were installed between the platform base plate and the test bench. During Y direction vibration, the vibration may cause preload loss in the fastening screws. This may be the main reason for the discrepancy between the experimental and simulation curves in the Y direction;

2) Data acquisition: As shown in Figures 16 and 17, the vibration response varies across different locations. The inconsistency between the positions of acceleration measurement points and simulation sampling nodes may result in discrepancies in acceleration amplitude. Additionally, the experimental curves were sampled at a variable rate, whereas the simulation curves maintained a fixed sampling frequency, which may also cause differences in the results.;

Combining the above analysis results, the actual fundamental frequency of the test platform is 436.2Hz, which meets the design value>300Hz indicator requirement, and leaves a certain design margin to meet the mechanical environment test needs of the coronagraph scientific probe module.

## 6 Conclusion

The research object of this paper is the mechanical test platform of the scientific probe module of the CPI-C. According to the module layout requirements and stiffness requirements, the test platform is optimized by response surface design. The fundamental frequency and amplitude response surface model are obtained by least squares fitting, and the final configuration parameters of the test platform are determined. The effectiveness of the optimization model is verified by dynamic simulation and sine frequency sweep test. The fundamental frequency of the test platform is 436.2Hz, which meets the mechanical test requirements of the scientific probe module of the CPI-C at all stages. At the same time, this design method can guide the structural design of the CPI-C and provide a theoretical basis for the implementation of the Chinese space station project.

## Acknowledgments

This research was supported by the National Natural Science Foundation of China (grants nos.11827804, U2031210 and 12103073) and the China Manned Space Project (grants nos. CMS-CSST-2025-A18, CMS-CSST-2021-A11, CMS-CSST-2021-B04, and CMS-CSST-201906)

## Data availability statements

The data that support the findings of this study are available from the corresponding author, upon reasonable request.

## Conflict of Interest Statement

The authors declare that they have no known competing financial interests or personal relationships that could have appeared to influence the work reported in this paper.